\documentclass[twocolumn,showpacs,preprintnumbers,amsmath,amssymb]{revtex4}
\usepackage{graphicx}
\usepackage{dcolumn}% Align table columns on decimal point
\usepackage{bm}
\usepackage{amssymb}
\usepackage{amsmath}
\usepackage{amsthm}
\usepackage{latexsym}
\newcommand{\be}{\begin{equation}}
\newcommand{\ee}{\end{equation}}
\newcommand{\bea}{\begin{eqnarray}}
\newcommand{\nn}{\nonumber}
\newcommand{\eea}{\end{eqnarray}}

\begin{document}
\title{Unification of inflation and cosmic acceleration in the Palatini formalism}

\author{Thomas P.~Sotiriou}
 \email{sotiriou@sissa.it}
\affiliation{SISSA-International School of Advanced Studies, via Beirut 2-4, 34014, Trieste, Italy.}

\begin{abstract}
A modified model of gravity with additional positive and negative 
powers of the scalar curvature, $R$, in the gravitational action is studied. This is done using the Palatini
variational principle. It is demonstrated that using such a model might prove useful to explain
both the early time inflation and the late time cosmic acceleration without the need for any form 
of dark energy.

\end{abstract}


\maketitle

%%%%%%%%%%%%%%%%%%%%%%
%%%%%%%%%%%%%%%%%%%%%%
\section{\label{sec:1}Introduction}
%%%%%%%%%%%%%%%%%%%%%%
It seems to be  a well established fact that the universe is undergoing a phase of accelerated expansion 
\cite{super,super2,torny,bennett,boom,halverson}. This is usually explained via the presence of 
the so-called dark energy \cite{peebles,padm}. This is not, however, the only way to address this 
problem. Several authors are suggesting that gravity should in some way be modified. One possible modification is to 
generalize the Einstein-Hilbert action by including higher-order terms in the scalar curvature (for a 
generic study see \cite{buh}). 
Such terms seem to be predicted, among other curvature invariants or non-minimally coupled scalar fields, in the effective action in almost every unification scheme, such as Superstrings, Supergravity or 
GUTs. For example, terms inversely proportional to $R$ may be expected from some 
time-dependent compactification of string or M theory \cite{noji2} and terms involving positive powers
of curvature invariants may be induced by the expansion of the effective action at large curvature \cite{vassi}.
  This of course does not mean that adding such higher order terms in the gravitational Lagrangian leads to a String/M theory motivated action. It is more of an indication of the fact that such phenomenological models seem too be just somehow inspired by these predictions. The choice of these generalized actions is, strictly speaking, based upon the resulting phenomenology. 

 In \cite{carroll} it was shown that a term inversely proportional to the curvature can lead to late time
expansion. Soon after concerns were raised on whether this model passes solar system tests \cite{chiba}
or has a correct Newtonian limit \cite{dick}. However, there are also serious doubts about whether these problems are 
really present or merely products of a misconception \cite{cem}. The most important drawback in adopting 
such a model still remains. It has to do with its stability in a weak gravity regime within matter \cite{dolgov} 
and might be overcome by more sophisticated models \cite{noji}.

In \cite{vollick} a further modification by using the Palatini variational principle on the action of \cite{carroll} was suggested. In
this formalism the metric and the affine connection are considered as independent quantities, and one
has to vary the action with respect to both of them in order to derive the field equations. If the
standard Einstein-Hilbert action is used, one will get standard general relativity with this approach \cite{wald},
as with the metric approach, but, for a more general action, the Palatini and the metric approaches 
give different results. The Palatini approach
seems appealing because the field equations are not fourth-order pde's as in the metric approach, but
 a set of second order pde's plus an equation involving the connection, which is trivial 
to solve and interpret using a conformal transformation \cite{vollick}. Additionally, in vacuum
the theory becomes equivalent to Einstein gravity plus a cosmological constant so it passes solar system tests as long as
this constant is chosen to have a suitable value. This also ensures that some interesting attributes of general relativity, 
like gravitational waves and black holes (Schwarzchild-de Sitter and anti-de Sitter) are present here as well.
Finally, whether models like the one discussed in \cite{vollick} have the correct Newtonian limit 
seemed to be a controversial issue \cite{meng2,barraco}, but a recent paper \cite{sot} seems to be
settling it, giving a positive answer.

The universe is also thought to have undergone an inflationary era at early times. The most common approach
is to assume that inflation is caused by a field manifested for exactly this purpose \cite{liddle}. There were suggestions 
in the past about modifying gravity in such a way as to explain inflation without the need of such a field \cite{staro}.
There were also attempts to unify inflation and late accelerated expansion in a single approach (see for example \cite{bo,odi}).
In \cite{noji} a model with both positive and negative powers of the scalar curvature in the action was
considered in the metric formalism in order to explain both the early time inflation and the late time
 cosmic expansion. In the Palatini formalism, the only similar attempt so far, but with only an extra $R^2$ term included in the action,
 was presented in \cite{meng}, and 
the authors derive the conclusion that this term cannot lead to gravity driven inflation. 

In the present paper the approach of \cite{meng} will be followed by using a Lagrangian with both positive and negative 
powers of $R$ in an attempt to explain both inflation and the present cosmic accelerated expansion. It will be shown that  
including an $R^3$ and an $R^{-1}$ term can account for both effects and lead to a unified model, whereas the $R^2$ term used in
\cite{meng} is ineffective. It is
important to note here that the scope of this paper is merely to demonstrate that it is possible, from a qualitative
point of view, to create such a unified model in the Palatini formalism. Thus possible problems with the Newtonian limit will not concern us here, although  it can easily be shown that the discussed model has the correct Newtonian behavior \cite{sot}.

The rest of the paper is organized as follows. In section \ref{sec:2} the Palatini formalism is 
very briefly revised. After that the discussed model is presented and studied in vacuum.
In section \ref{sec:3} FLRW cosmology is studied in this modified gravity regime. It is shown that it exhibits an 
inflationary behavior at early times, an era similar to standard cosmology, and a de Sitter expansion at 
late time. Section \ref{sec:4} contains conclusions.

%%%%%%%%%%%%%%%%%%%%%%
%%%%%%%%%%%%%%%%%%%%%%
\section{\label{sec:2}Palatini formalism}
%%%%%%%%%
%%%%%%%%%%%%%%%%%%%%%%
Let us, very briefly, review the Palatini formalism for a generalized action of the form
\be
S=\frac{1}{2\kappa^2}\int d^4 x\sqrt{-g}L(R)+S_M,
\ee
where $\kappa^2=8\pi G$ and $S_M$ is the matter action. For a detailed study see \cite{vollick}.

Varying with respect to the metric $g_{\mu\nu}$ gives
\be
\label{struct}
L'(R) R_{\mu\nu}-\frac{1}{2}L(R)g_{\mu\nu}=\kappa^2 T_{\mu\nu},
\ee
where the prime denotes differentiation with respect to $R$ and the stress-energy tensor (SET) $T_{\mu\nu}$
is given by
\be
T_{\mu\nu}=-\frac{2}{\sqrt{-g}}\frac{\delta {\cal L}_M}{\delta g^{\mu\nu}},
\ee
where ${\cal L}_M$ is the matter Lagrangian density.
Since the affine connection and the metric are considered to be geometrically independent objects we should
also vary the action with respect to $\Gamma^{\lambda}_{\mu\nu}$. If we also use the fact that the connection is
the Christoffel symbol of the conformal metric $h_{\mu\nu}\equiv L'(R) g_{\mu\nu}$ (see \cite{vollick}),
we get
\be
\Gamma^{\lambda}_{\mu\nu}= \left\{^{\lambda}_{\mu\nu}\right\} +\frac{1}{2 L'}
\left[2\delta^{\lambda}_{(\mu}\partial_{\nu)}L'-g_{\mu\nu}g^{\lambda\sigma}\partial_{\sigma}L'\right],
\ee
where $\left\{^{\lambda}_{\mu\nu}\right\}$ denotes the Christoffel symbol of $g_{\mu\nu}$.
By contracting eq. (\ref{struct}) one gets
\be
\label{scalar}
L'(R) R-2 L(R)=\kappa^2 T.
\ee
It is important to notice that, once one chooses a certain L(R), when $T=0$, as in vacuum for example (but not only in that case as we are going to see
later on), eq. (\ref{scalar}) becomes a simple
algebraic equation in $R$. This implies that $R$ is a constant and the studied theory becomes nothing more
than Einstein gravity plus a cosmological constant. Therefore in vacuum, the maximally symmetric solutions will be de Sitter and anti-de Sitter and flat spacetime will not be a global solution, as in the cosmological constant model. Consequently any solution will not be asymptotically flat but asymptotically de Sitter and anti-de Sitter. For example the static black hole solutions will be, instead of Schwarzchild, Schwarzchild-de Sitter and Schwarzchild-anti-de Sitter solutions.

In \cite{meng} the authors considered a modified action with an additional $R^2$ term and they arrived at
the result that this cannot lead to a gravity-driven inflation. However, including an $R^2$ term
seems to be a singular case, since as one can easily see from the form of eq. (\ref{scalar}), it will
not give any contribution to the left hand side. Here we will consider the Lagrangian
\be
\label{lagrang}
L(R)=\frac{R^3}{\beta^2}+R-\frac{\epsilon^2}{3 R},
\ee
where $\epsilon$ and $\beta$ are for the moment some constants, on which we will try to put constraints later on. Our choice of the form of the Lagrangian is based upon the interesting phenomenology that it will lead to.
In vacuum eq. (\ref{scalar}) gives for $L$ given by eq. (\ref{lagrang})
\be
\label{alg}
R^4-\beta^2 R^2+\epsilon^2\beta^2=0.
\ee
Note that even if we included an $R^2$ term in eq.~(\ref{lagrang}), eq.~(\ref{alg}) would remain
unchanged due to the form of eq. (\ref{scalar}). Thus, even if we have avoided that for simplicity's
sake, there is no reason to believe that it will seriously affect our results in any way. One can easily
 solve eq.~(\ref{alg}) to get
\be
R^2=\frac{\beta^2}{2} \left[1\pm\sqrt{1-4\left(\epsilon/\beta\right)^2}\right].
\ee
If $\epsilon\ll\beta$  this corresponds to two de Sitter and two
anti-de Sitter solutions for $R$. Here we will consider the two de Sitter solutions, namely:
\be
\label{solutions}
R_1\sim \beta,\quad R_2\sim \epsilon.
\ee
If we further assume that $\epsilon$ is sufficiently small and $\beta$ is sufficiently large, then, since
the expansion rate of the de Sitter universe scales like the square root of the scalar curvature, 
$R_1$ can act as the seed of an early time inflation and $R_2$ as the seed of a late time accelerated
expansion.
%%%%%%%%%%%%%%%%%%%%
%%%%%%%%%%%%%%%%%%%%
\section{\label{sec:3}FLRW Cosmology}
%%%%%%%
%%%%%%%%%%%%%%%%%%%%
Let us now check our previous result in Friedmann-Lemaitre-Robertson-Walker cosmology. We are going to
consider the spatially flat metric
\be
ds^2=-dt^2+a(t)^2 (dx^2+dy^2+dz^2),
\ee
which is favored by present observations (see for example \cite{bennett,boom}). We also assume that the stress-energy tensor
is that of a perfect fluid, i.e.
\be
T_{\mu\nu}=(\rho+p)u_{\mu}u_{\nu}+p\, g_{\mu\nu}.
\ee
Following \cite{meng} we write the non-vanishing
components of the Ricci tensor:
\be
\label{ricci1}
R_{00}=-3\frac{\ddot{a}}{a}+\frac{3}{2}(L')^{-2}(\partial_{0} L')^2-\frac{3}{2}(L')^{-1}\bar{\nabla}_0\bar{\nabla}_0 L',
\ee
\bea
\label{ricci2}
R_{ij}=[a \ddot{a}+2 \dot{a}^2 &+& (L')^{-1}\left\{^{\lambda}_{\mu\nu}\right\} \partial_0 L'
+{}\nn\\& &{}
+\frac{a^2}{2}(L')^{-1}
\bar{\nabla}_0\bar{\nabla}_0 L']\delta_{ij},
\eea
where $\bar{\nabla}$ is the covariant derivative associated with $g_{\mu\nu}$.
Using eqs. (\ref{struct}), (\ref{ricci1}) and (\ref{ricci2}) we can derive the modified Friedmann
equation:
\bea
\label{mf}
6 H^2&+&6H(L')^{-1}\partial_0 L'+\nn\\
& &+\frac{3}{2}(L')^{-2}(\partial_0 L')^2=\frac{\kappa^2(\rho+3 p)+L}{L'}.
\eea
where $H\equiv \dot{a}/a$ is the Hubble parameter.

Now we can investigate the different cosmological eras of our model. At very early time we expect
the matter to be fully relativistic. Denoting by $\rho_r$ and $p_r$ the energy density and pressure, the equation of state will be $p_r=\rho_r /3$. Thus $T=0$ and 
eq. (\ref{scalar}) will have the solution we mentioned in the previous section. Taking into account
that we expect the curvature to be large we infer that $R=R_1=\beta$. Therefore, the universe will
undergo a de Sitter phase which can account for the early time inflation. As usual, conservation of 
energy implies $\rho_r \sim a^{-4}$.
Thus, it is easy to verify that the last term on the right hand side of eq. (\ref{mf}) will quickly 
dominate. Additionally, $R$ is a constant now so the second and third term on the left hand side vanish and 
we are left with
\be
H\sim\sqrt{\beta/12}.
\ee

At some stage the temperature will drop enough for some matter component to become non-relativistic \cite{note}. If we assume that its pressure $p_m=0$
(dust), then eq. (\ref{scalar}) takes the following form:
\be
\label{eq1}
\frac{R^3}{\beta^2}-R+\frac{\epsilon^2}{R}=-\kappa^2\rho_m.
\ee
SET conservation now implies
\be
\label{eq2}
\dot{\rho}_m+3H\rho_m=0.
\ee
Using eqs. (\ref{eq1}) and (\ref{eq2}) it is easy to show after some mathematical manipulations
 that
\be
\label{eq3}
\dot{R}=\frac{3H R\left(R^2-\frac{R^4}{\beta^2}-\epsilon^2\right)}{\left(\frac{3 R^4}{\beta^2}-R^2-\epsilon^2\right)}.
\ee
The modified Friedmann equation (\ref{mf}) takes the form
\be
\label{mf2}
H^2=\frac{ 2\kappa^2 \rho+\Lambda}
{6\left(\frac{3 R^2}{\beta^2}+1+\frac{\epsilon^2}{3 R^2}\right) \left(1+\frac{3}{2} A\right)^2}
\ee
where $\rho=\rho_r+\rho_m$, 
\bea
A&=&\frac{\left(\frac{6 R^4}{\beta^2}-\frac{2}{3}\epsilon^2\right) \left( R^2-\frac{R^4}{\beta^2}-\epsilon^2\right)}
{\left(\frac{3 R^4}{\beta^2}-R^2-\epsilon^2\right)\left(\frac{3 R^4}{\beta^2}+R^2+\frac{\epsilon^2}{3}\right)},\\
\Lambda&=&2\left(\frac{R^3}{\beta^2}+\frac{\epsilon^2}{3R}\right),
\eea
and we have used eq. (\ref{eq1}) and the equation of state for the relativistic component of the cosmological fluid.

\begin{figure}[t]
\begin{center}
\includegraphics[width=6cm,angle=0]{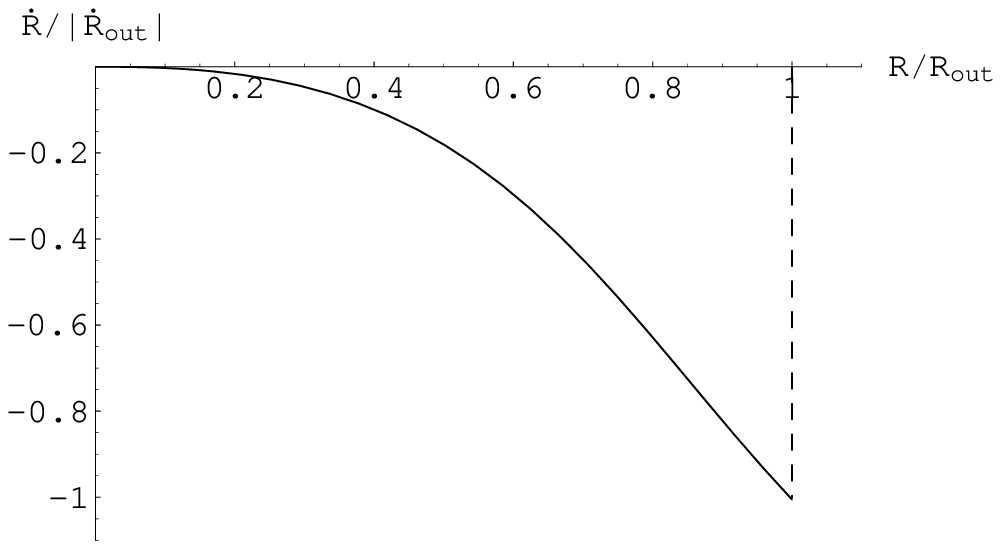}
\caption{$\dot{R}$ as a function of $R$.}\label{fig1}\includegraphics[width=6cm,angle=0]{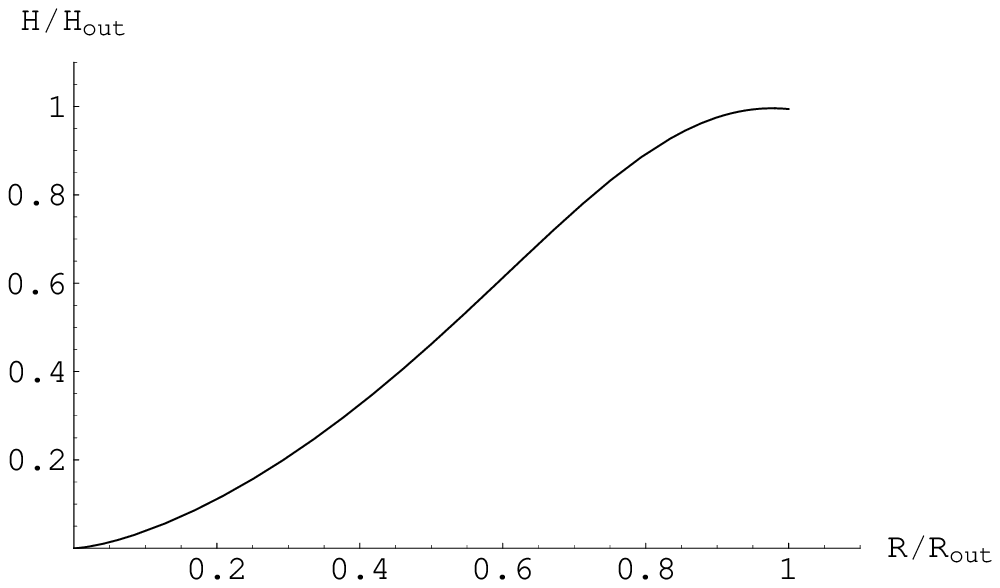}
\caption{$H$ as a function of $R$. It is easy to see from these figures that,  in the presence of non-relativistic matter, the universe will start to become flatter and the expansion rate will drop significantly.
 The subscript ``$out$'' denotes, in both graphs, the value of the quantities at
the time when the pure de Sitter phase ends.}\label{fig2}
\end{center}
\end{figure}

 Let us now use the equations we derived to examine the behavior of the universe in lower curvature. The second term in the numerator on the right hand side of eq. (\ref{mf2}), $\Lambda$, is
dominant with respect to the matter term as long as the curvature is still very large. Thus we can use
eqs. (\ref{mf2}) and (\ref{eq3}) to derive $\dot{R}$ as a function only of $R$, i.e.
\be
\dot{R}=f(R).
\ee
Due to its length we will not give the expression for $f$ explicitly. We can, however,
understand its behavior, by plotting its graph (see Fig.~\ref{fig1}).
One can easily see that  in the presence of a non-relativistic matter component the curvature will
start to decrease dramatically. The same holds for the Hubble parameter $H$ (Fig.~\ref{fig2}). Therefore, in a short time the curvature will become small enough so that the matter term will dominate eq. (\ref{mf2}). At the same time $\dot{R}$ will also become very small. The above can also be inferred in the following
way. $\Lambda$ plays the role of an effective dynamical cosmological constant. Since the curvature is 
now less than $\beta$ and as long as it is significantly larger than $\epsilon$, in the leading order
\be
\dot{R}\sim -3 H R,\qquad H^2\sim \frac{R^3}{3\beta^2}
\ee
Thus, it is easy to see that
\be
\label{r1}
R\sim t^{-2/3},\qquad a(t)\sim t^{2/9}.
\ee
From eqs. (\ref{r1}) one concludes that
\be
\rho_r\sim t^{-8/9},\qquad
\Lambda\sim t^{-2}.
\ee
i.e.~$\Lambda$ decreases much faster than the energy density of relativistic matter. Hence, the 
universe will soon enter a radiation dominated era characterized by a very low curvature. 

We next investigate the behavior of the modified Friedmann equation (\ref{mf}) for low curvature.
The second and the third term on the right hand side are negligible now and 
$L'=3R^2/\beta^2+1+\epsilon^2/(3R^2)$ will tend to $1$ provided that $\epsilon$ is small enough. On the
other hand, $L=R^3/\beta^2+R-\epsilon^2/(3R)$ can be written, using eq. (\ref{eq1}), as
\be
L=\frac{2 R^3}{\beta^2}+\frac{2\epsilon^2}{3 R} +\kappa^2\rho_m,
\ee 
where the dominant term now is $\kappa^2 \rho_m$. Therefore, eq. (\ref{mf}) will give
\be
\label{stand}
H^2\sim \kappa^2\rho,
\ee
where, as before, $\rho=\rho_r+\rho_m$. Eq. (\ref{stand}) resembles standard cosmology. It is reasonable, therefore to assume that everything can continue
as expected, i.e. radiation dominated era, Big Bang Nucleosynthesis (BNN), and matter dominated era. Recall that the scenario 
described above starts at very high curvature and as soon as matter starts becoming non-relativistic 
so it seems safe to assume that radiation will still be dominant with respect to ordinary matter (dust).

Finally at some point we expect matter to become subdominant again due to the increase of the scale factor.
Thus at late times we can arrive to the picture where $\rho\sim 0$ and eq. (\ref{scalar})
will give the vacuum solutions mentioned in the previous section (eq.~(\ref{solutions})). However, now the curvature is small and the preferred solution
will be $R_2\sim \epsilon$. The universe will therefore again enter a de Sitter phase of accelerated expansion 
similar to the one indicated by current observations. 

The values of $\epsilon$ and $\beta$ are left unspecified here, apart from some loose assumptions 
about their order of magnitude and can be fixed to match observations. Specifically, if we assume that the curvature
is already of the order of its asymptotic value, then $\epsilon$ should be
of the order of $10^{-67} (\textrm{eV})^2$ to account for the current accelerated expansion. On the other hand, 
the value of $\beta$ is related to  the expansion rate during inflation.
So, probably some bounds on its value will be put
by more carefully examining the inflationary behavior of such a model. Solving the flatness seems to be trivial. However, one will need a significant
number of e-foldings to address the horizon problem and create the perturbations in the Cosmic Microwave Background and the value of $\beta$ might have a significant role on that. As a first comment let us also say that having $\beta$ very large and $\epsilon$ very small is actually the most physical choice, because it represents a small deviation away from general relativity.

%%%%%%%%%%%%%%%%%%%%%%%%%%%%%%%%%
%%%%%%%%%%%%%%%%%%%%%%%%%%%%%%%%%
\section{\label{sec:4}Conclusions}
%%%%%%%%%%%%%%%%%%%%
%%%%%%%%%%%%%%%%%%%%%%%%%%%%%%%%%

A model of modified gravity including both positive and negative powers of the scalar curvature
in the action has been
considered here within the Palatini approach. It has been demonstrated that such a model
can account, apart from late time accelerated cosmic expansion which was a well established fact, also for early time inflation.  At the same time its evolution during the standard cosmological eras, like Big Bang Nucleosynthesis is almost identical to the standard cosmological model.
 
 It should be noted, of course, that the approach is highly qualitative. 
A detail quantitative approach should be performed to check the conclusions of this study, 
whose scope was merely to demonstrate that it seems possible
to create a unified model for inflation and cosmic expansion in the Palatini formalism. 
 Inflation is needed to solve specific problems related with the cosmological evolution, like the generation of large enough density perturbations and the Horizon problem. Therefore one would like the model to provide us with an inflationary behavior that can address these problems successfully and be in agreement with the observations. 
Another point that has to be studied further is the mechanism that provides the passage from inflation to ordinary cosmology. It is not clear at the moment how such a geometrical inflation will end and how the universe will reheat after that. We will address this point in future work \cite{sot2}.

%%%%%%%%%%%%%%%%%%%%%%%%%%%%%
\section*{Acknowledgements}
The author would like to thank Stefano Liberati, John Miller and Francesca Perrotta for drawing his attention to 
this project, and for valuable discussions and comments.
%This research was supported in part by Grant No 70/4/4056 of the Special Account for Research Grants of the
%University of Athens, and in part by Grant No 70/3/7396 of the ``PYTHAGORAS'' research funding program.


\begin{thebibliography}{20}
\bibitem{note} Note that this does not imply that $\rho_m>\rho_r$. We can still be in regime that $\rho_r\gg\rho_m$ and
the heaviest species of particles, for example, start to become non-relativistic. We have no radiation term in the
right hand side of eq. (\ref{eq1}) just because $p_r=\rho_r /3$.
\end{thebibliography}
\end{document}